\documentclass[prl, a4paper, twocolumn, 10pt, showpacs]{revtex4}
\usepackage{bbm, amsmath, amssymb, amsthm, bm,textcomp, nicefrac}
\usepackage[T1]{fontenc}
\usepackage[latin9]{inputenc}
\usepackage{graphicx}
\usepackage{geometry}

\newcommand{\ket}[1]{\left|{#1}\right\rangle}

\newcommand{\ketbrad}[1]{\left|{#1}\rangle\!\langle{#1}\right|}
\newcommand{\ketbra}[2]{\left|{#1}\rangle\!\langle{#2}\right|}

\newcommand{\be}{\begin{equation}}
\newcommand{\ee}{\end{equation}}
\newcommand{\eea}{\end{eqnarray}}
\newcommand{\bea}{\begin{eqnarray}}
\newcommand{\Def}{ \mathrel{\mathop:}=}

\begin{document}

\title{Are cloned quantum states macroscopic?}

\author{F.\ Fr\"owis and W.\ D\"ur}

\affiliation{Institut f\"ur Theoretische Physik, Universit\"at
  Innsbruck, Technikerstra\ss e 25, A-6020 Innsbruck,  Austria}
\date{\today}

\begin{abstract}
We study quantum states produced by optimal phase covariant quantum cloners. We argue that cloned quantum superpositions are not macroscopic superpositions in the spirit of Schr\"odinger's cat, despite their large particle number. This is indicated by calculating several measures for macroscopic superpositions from the literature, as well as by investigating the distinguishability of the two superposed cloned states. The latter rapidly diminishes when considering imperfect detectors or noisy states, and does not increase with the system size. In contrast, we find that cloned quantum states themselves are macroscopic, in the sense of both proposed measures and their usefulness in quantum metrology with an optimal scaling in system size. We investigate the applicability of cloned states for parameter estimation in the presence of different kinds of noise.
\end{abstract}




\maketitle


Quantum mechanics predicts puzzling features of quantum systems, mainly due to the possibility of superposition and interference. While such properties are regularly observed at a microscopic scale, the production of large-scale entanglement or superpositions of macroscopic objects has become the focus of interest in recent years. Remarkable experimental progress has been reported with various setups, including atomic systems \cite{ExperimentsAtoms}, superconducting devices \cite{ExpSuperConduct,Leg02}, and optomechanical systems \cite{ExpOptoMech} as well as photons \cite{DeMartini08,De12}. Some of these experiments aim for the production of a quantum superposition of macroscopic objects, in the spirit of the famous Schr\"odinger cat gedankenexperiment \cite{Schroedinger35}.

However, the required features that make a quantum object macroscopic or a macroscopic superposition are not obvious. Should one call a single atom in a superposition of two localized states macroscopic if the two wavepackets are far apart? What about massive objects, or objects consisting of large number of particles, where the displacement of the two wavepackets is extremely small?
Several attempts have been put forward in the literature that aim for providing measures of macroscopicity and macroscopic superpositions \cite{Leg80,Du02,p-Index,BM,KWDC,MvD,LJ,FD12}. Here we will concentrate on discrete systems consisting of $N$ qubits, where the notion of macroscopicity is better explored \cite{FD12}. In this case it is evident that a large particle number $N$ is required to call such a state macroscopic; however, this is apparently not sufficient. For instance, a product state of $N$ qubits is certainly not macroscopically entangled, and also a quantum state where two particles are in a superposition but the rest is in some fixed product state is not a macroscopic superposition either.

We consider quantum states produced by an optimal phase covariant quantum cloner (PCQC) \cite{PCQC}. These states have been considered in the context of photons, where it has been demonstrated that an optimal PCQC can be experimentally realized by using parametric amplification \cite{ExpRealizationPCQC}, leading to quantum states consisting of tens of thousand photons \cite{DeMartini08,De12,EyeDetector,Po11}. When applied to an initial state in a superposition, or part of a maximally entangled photon pair, this leads to quantum superpositions of two such multiphoton states. It has been argued that with such states it is possible to observe entanglement with the naked eye \cite{EyeDetector} and that indeed this corresponds to the production of micro-macro entanglement in the spirit of Schr\"odinger's cat \cite{DeMartini08,De12}.

Here, we critically investigate the macroscopicity of such systems. Thereby, we carefully distinguish between macroscopicity of the states as such and macroscopic superpositions \cite{FD12}. By applying different measures from the literature \cite{BM,KWDC,MvD,FD12} that have been proposed to determine an effective size of quantum superpositions, we show that all these measures render cloned superpositions microscopic, despite their large particle number. This conclusion is also supported by considering the distinguishability of the two superposed states, which rapidly diminishes when considering noisy states or imperfect detectors (see also Refs.~\cite{EyeDetector,NoisyDetectors}) and which does not increase with the number of particles. In contrast, cloned states as such turn out to be macroscopic, both in the sense of other proposed measures \cite{p-Index,FD12}, but also with respect to their applicability for parameter estimation with enhanced precision, since the Heisenberg limit can be reached \cite{Covariance}. We investigate the applicability of cloned states for quantum metrology \cite{Giovanetti} in the presence of different kinds of noise and imperfections.

\subsection{Phase covariant quantum cloner}

We start by reviewing the optimal PCQC \cite{PCQC}. We consider systems consisting of $N$ qubits with Hilbert space ${\cal H}= (\mathbbm{C}^2)^{\otimes N}$. A PCQC produces from a single qubit state  $|\varphi\rangle=\frac{1}{\sqrt{2}}(|0\rangle + e^{i \varphi}|1\rangle)$ lying on the equatorial plane of the Bloch sphere $N$ optimal copies, in the sense that the resulting symmetric $N$-qubit state is such that the fidelity of each single particle reduced density operator with respect to the initial state is maximal. The optimal unitary cloning operation performs for odd $N$ the mapping \cite{PCQC}
\begin{equation}
|\pm\rangle \rightarrow |\psi^{\pm}\rangle \equiv \frac{1}{\sqrt{2}}\left(\left|N,\frac{N-1}{2}\right\rangle \pm \left|N,\frac{N+1}{2}\right\rangle\!\!\right),\label{eq:3}
\end{equation}
where $|\pm\rangle = \frac{1}{\sqrt{2}}(|0\rangle \pm |1\rangle)$ are the eigenstates of $\sigma_x$, and $|N,k\rangle$ denotes a $N$-qubit symmetric Dicke state with $k$ excitations \cite{Dicke}, that is, a state given by a coherent superposition of all possible permutations of product states with $k$ ones and $N-k$ zeros. Note that, for even $N$, Eq.~(\ref{eq:3}) differs in the details. However, the results and conclusions are equivalent to odd $N$, on which we focus for the remainder of this article.

Applying the cloning transformation to the second qubit of an entangled state $\frac{1}{\sqrt{2}}(|+\rangle\otimes|-\rangle - |-\rangle\otimes|+\rangle)$ results in an $N+1$ qubit state
\begin{equation}
\label{eq:1}
|\psi\rangle = \frac{1}{\sqrt{2}}(|+\rangle\otimes |\psi^{-}\rangle - |-\rangle\otimes|\psi^{+}\rangle).
\end{equation}
This is the state we will consider in the following.
Notice that it was argued in Refs.~\cite{De12,ExpRealizationPCQC,SimonOptCloning} that Eq.~(\ref{eq:3}) corresponds to the quantum state that is generated in a photonic setup using parametric amplification.

\subsection{Measures for macroscopic superpositions}
We now review and apply some measures to quantify the ``effective size'' $N_{\mathrm{eff}}$ of multipartite spin states of the form
\begin{equation}
\label{eq:2}
\left| \phi  \right\rangle = \frac{1}{\sqrt{2}}\left( \left| \phi_0 \right\rangle + \left| \phi_1 \right\rangle  \right).
\end{equation}
For the state (\ref{eq:1}), we use the splitting $\left| \phi_0 \right\rangle = |+\rangle\otimes |\psi^{-}\rangle$ and $\left| \phi_1 \right\rangle = |-\rangle\otimes |\psi^{+}\rangle$.

The first two measures try to capture the notion of ``macroscopically distinct'' states $\left| \phi_0 \right\rangle $ and $\left| \phi_1 \right\rangle $. Korsbakken \textit{et al.} \cite{KWDC} argue that these two states are macroscopically distinct if they are distinct already on a microscopic subset of the whole number of particles. Hence the effective size  $N_{\mathrm{eff}}^{\mathrm{K}}$ of a quantum state (\ref{eq:2}) is defined as the maximal number of partitions among the particles such that measuring just one of these groups allows us to differentiate between $\left| \phi_0 \right\rangle $ and $\left| \phi_1 \right\rangle $ with high probability.

\begin{figure}[ht]
 \includegraphics[width=\columnwidth]{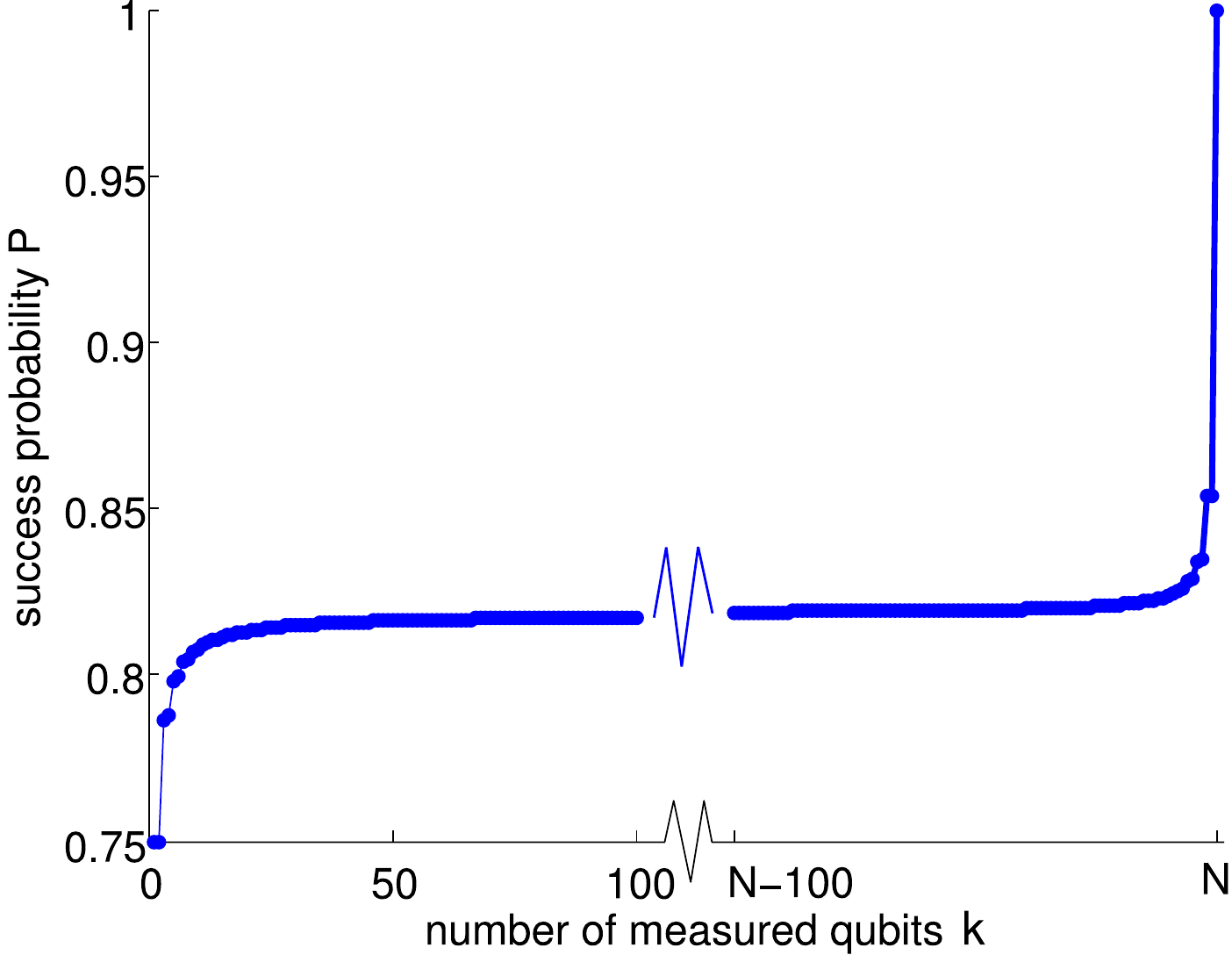}
  \caption[]{\label{fig:korsbakken} 
    Success probability $P$ to distinguish $\left| \psi^{+} \right\rangle $ from $\left| \psi^{-} \right\rangle $ by measuring $k$ out of $N$ particles. The results are valid in the limit of large $N$. We clearly see that there is little difference between the scenarios where one measures 100 particles and \textit{all but} 100 particles, for which we have in both cases $P \approx 82 \%$.  }
\end{figure}

To apply the measure we first note that the ``micropart'' in (\ref{eq:1}) constitutes one group which allows with a $\sigma_x$ measurement a perfect distinction between $\left| \phi_0 \right\rangle$ and $\left| \phi_1 \right\rangle$. The nontrivial part is to determine the optimal division of the cloner states with respect to a success probability $P$ for their distinction. This probability is defined as $P \Def \frac{1}{2} + \frac{1}{4} \lVert \rho_k^+ - \rho_k^{-} \rVert_1$ \cite{Note2} with $\rho_k^{\pm} = \mathrm{Tr}_{N \setminus k}\left| \psi^{\pm} \rangle\!\langle \psi^{\pm}\right| $. The detailed calculation is presented in the Appendix (see also Ref.~\cite{De12}). For $k=1$, we have $P= \frac{3}{4}[1+O(1/N)]$. If we increase $k$, the success probability goes up, but saturates to a value below 0.82; see Fig.~\ref{fig:korsbakken}. On the other side, we can also measure \textit{all but} $k$ particles. From Fig.~\ref{fig:korsbakken}, we see that it makes little difference whether we measure $100$ or $N-100$ particles for $N\gg 1$. Even if we measure $N-1$ particles, $P \approx \frac{1}{2}(1+\frac{1}{\sqrt{2}}) \approx 0.85$. The effective size depends on the uncertainty in the measurements we are willing to allow. Nevertheless, it seems clear that we have to measure \textit{all} particles to be certain whether we encounter $\ket{\psi^{+}}$ or $\ket{\psi^{-}}$. Hence the state is not a macroscopic superposition; that is, $N_{\mathrm{eff}}^{\mathrm{K}}(\psi) = 2$.

Similarly, Marquardt and co-workers \cite{MvD} see two states distinct on a macroscopic scale if one has to apply \textit{on average} $O(N)$ one-particle operators on $\left| \phi_0 \right\rangle $ in order to achieve a high overlap with $\left| \phi_1 \right\rangle$. For the state $\left| \psi \right\rangle $ of Eq.~(\ref{eq:1}), we need at least two one-particle operations because the microscopic as well as the macroscopic parts are orthogonal on their own. But we will see that we also do not need more than two; therefore $N_{\mathrm{eff}}^{\mathrm{M}}(\psi) = 2$, which means that the superposition is not macroscopic in this sense. We prove the statement with the operator $\sigma_z\otimes M_z$, where $M_j = \sum_{i = 1}^N \sigma_j^{(i)}, j = x,y,z$ \cite{NoteNotation}. The Dicke states are eigenstates of $M_z$ with $M_z \left| N,\frac{1}{2}(N\pm 1) \right\rangle  = \mp \left| N,\frac{1}{2}(N\pm 1) \right\rangle$, which means that $\sigma_z\otimes M_z \left| \phi_0 \right\rangle  = \left| \phi_1 \right\rangle$. With $\sigma_z\otimes M_z$ as a linear combination of two-particle operations, we entirely map $\left| \phi_0 \right\rangle $ onto $\left| \phi_1 \right\rangle $ and vice versa.

Bj\"ork and Mana \cite{BM} call a superposition of the type (\ref{eq:2}) macroscopic if the usefulness of $\left| \phi \right\rangle $ is largely increased compared to the single constituents $\left| \phi_0 \right\rangle $ and $\left| \phi_1 \right\rangle $. This usefulness was defined in terms of interferometry experiments. For this it suffices that a local Hamiltonian $H$ generates a unitary evolution which is much more rapid for $\left| \phi \right\rangle $ than for $\left| \phi_0 \right\rangle $ and $\left| \phi_1 \right\rangle $. The ``speed of evolution'' is measured by the time $\theta_{\bot}$ it takes to evolve the state to
an orthogonal one. The effective size of the state (\ref{eq:2}) is hence the ratio $N_{\mathrm{eff}}^{\mathrm{B}}(\phi)=\theta_{\bot}(\phi_0)/\theta_{\bot}(\phi)$ \cite{NoteBM}. Since for spin systems $N_{\mathrm{eff}}^{\mathrm{B}}$ is not always defined (e.g., when $\left| \phi_0 \right\rangle $ is an eigenstate of $H$), we suggest to use instead the ``relative Fisher information'' \cite{FD12}, which is based on the quantum Fisher information $\mathcal{F}(\rho,H)$ \cite{Fisher} (see also the last part of the Appendix). Note that for pure states $\left| \phi \right\rangle $, $\mathcal{F}(\phi,H)$ is proportional to the variance of $H$, that is, $\mathcal{F}(\phi,H) = 4\mathcal{V}_{\phi}(H)$. To measure the usefulness of $\left| \phi \right\rangle $ compared to its constituents in Eq.~(\ref{eq:2}), we set $N_{\mathrm{eff}}^{\mathrm{rF}}(\phi) = \max_{H:local}\mathcal{V}_{\phi}(H)/\left[ \max_{H_0:local} \mathcal{V}_{\phi_0}(H_0) \right]$. If $N_{\mathrm{eff}}^{\mathrm{rF}}(\phi) = O(N)$, we call $\left| \phi \right\rangle $ a macroscopic superposition due to Bj\"ork and Mana; a constant scaling signifies a ``microscopic'' superposition.
We hence calculate the maximal variance of $\left|\psi  \right\rangle $ and $\left| \phi_0 \right\rangle $ with respect to all sums of local operators with a constant operator norm. For Dicke states with $N/2$ excitations this has already been done in Ref.~\cite{Covariance}, by using the so-called covariance method. For the micro-macro state $\left| \psi \right\rangle $ this can be easily adapted (for details, see the Appendix). The optimal operators yield $\mathcal{V}_{\psi}(H^{\mathrm{opt}}) = \frac{1}{2}(N+1)^2 + N +1$ and $\mathcal{V}_{\phi_0}(H_{0}^{\mathrm{opt}}) = \frac{1}{2}(N+1)^2$. The ratio of those two is the relative Fisher information $N_{\mathrm{eff}}^{\mathrm{rF}}(\psi) = 1 + 2/(N+1)$. We see that the quantum state is a microscopic superposition, because the two constituting states $\left| \phi_0 \right\rangle $ and $\left| \phi_1 \right\rangle $ are very similar to the total state $\left| \psi \right\rangle $ with respect to the maximal variances.

\subsection{Distinguishability of cloned superpositions}
In Refs.~\cite{DeMartini08,De12,Wigner} it has been argued that superposition (\ref{eq:1}) is macroscopic since the two macro states $\left| \psi^{+} \right\rangle $ and $\left| \psi^{-} \right\rangle $ are ``macroscopically distinguishable.'' This characterization is footed on the difference of the expectation values for the magnetization in the $x$ direction, which scales linearly with $N$, since $\langle M_x \rangle_{\psi^{\pm}} = \pm \frac{1}{2}N$. This reasoning implicitly assumes that the two states $\left| \psi^{\pm} \right\rangle $ are better and better distinguishable as $N$ grows, especially if noise or imperfections of the measurements are taken into account.

Here we show that this is not the case (see also Refs.~\cite{EyeDetector,NoisyDetectors}). If we assume noiseless measurements, the two states are perfectly distinguishable with respect to $M_x$ for all $N$, because $\left| \psi^{\pm} \right\rangle $ live exclusively in subspaces of $M_x$ with an even and odd excitation number, respectively. On the other side, the variances of $\left| \psi^{\pm} \right\rangle $ with respect to $M_x$ scale quadratically with $N$, $\mathcal{V}_{\psi^{\pm}}(M_x) = \frac{1}{4}\left( N^2-1 \right)$, which hinders an arbitrarily high distinguishability in the presence of noise even in the limit of large $N$. Given the spectral decomposition of a measurement $M = \sum_i m_i \pi_i$ with the eigenvalues $m_i$ and the (possibly high-dimensional) projections $\pi_i$, the quantity $D = \frac{1}{2} + \frac{1}{4}\Delta$ with $\Delta = \sum_i \lvert \langle \pi_i \rangle_{\psi^{+}} - \langle \pi_i \rangle_{\psi^{-}} \rvert$ serves as a characterization of the distinguishability between $\left| \psi^{+} \right\rangle $ and $\left| \psi^{-} \right\rangle $.

Here, we consider three different scenarios for $M$. The first scenario is a toy model which mimics a nonperfect resolution of a measurement apparatus. We therefore group the $N+1$ eigenvalues of $M_x$ into pairs, such that $m_i, m_{i+1} \rightarrow \frac{1}{2}\left( m_i+m_{i+1} \right)$ if $i$ is odd. This means that we cannot distinguish between even and odd parity of the $M_x$ measurement. Hence, one has $\Delta^{\mathrm{Pair}} =  \sum_{i= 1,3,\dots} \lvert \langle \pi_i + \pi_{i+1} \rangle_{\psi^{+}} - \langle \pi_i + \pi_{i+1}\rangle_{\psi^{-}} \rvert$. The second approach is to regard a generalized positive operator-valued measure (POVM). The measurement outcomes $m_i$ are associated with the Kraus operators $E_i = \sum_{j = 0}^N \sqrt{n_j} e^{-(i-j)^2/(4\sigma^2)} \pi_j$ where $n_j = \sum_{k = 0}^N e^{-(k-j)^2/(2\sigma^2)}$ is the normalization of the POVM. The parameter $\sigma > 0$ reflects the normally distributed uncertainty induced by the measurement. Hence we study $\Delta^{\mathrm{POVM}} = \sum_i \lvert \langle E^2_i \rangle_{\psi^{+}} - \langle E^2_i \rangle_{\psi^{-}} \rvert$. Here we choose $\sigma = \sqrt{N}$, but the conclusions are the same for constant $\sigma \gtrsim 1$. Last, we investigate $D$ if the states under consideration are subject to noise. We study a local phase-noise channel $\mathcal{E}_z$. This map acts on an arbitrary input state $\rho$ as $\mathcal{E}_z(\rho) = \prod_{i=1}^{N} u \rho + (1-u) \sigma_z \rho \sigma_z$ with the noise parameter $u = \frac{1}{2}\left( 1+e^{-\gamma t} \right) \in [\frac{1}{2},1]$. The corresponding $\Delta^{\mathrm{Noise}}$ equals $\sum_i \lvert  \mathrm{Tr}\left[\pi_i \mathcal{E}_z\left(\left| \psi^{+} \rangle\!\langle\psi^{+} \right| \right)\right]  -  \mathrm{Tr}\left[ \pi_i \mathcal{E}_z\left(\left| \psi^{-} \rangle\!\langle\psi^{-} \right| \right)\right] \rvert$.

\begin{figure}[htbp]
\centerline{\includegraphics[width=.9\columnwidth]{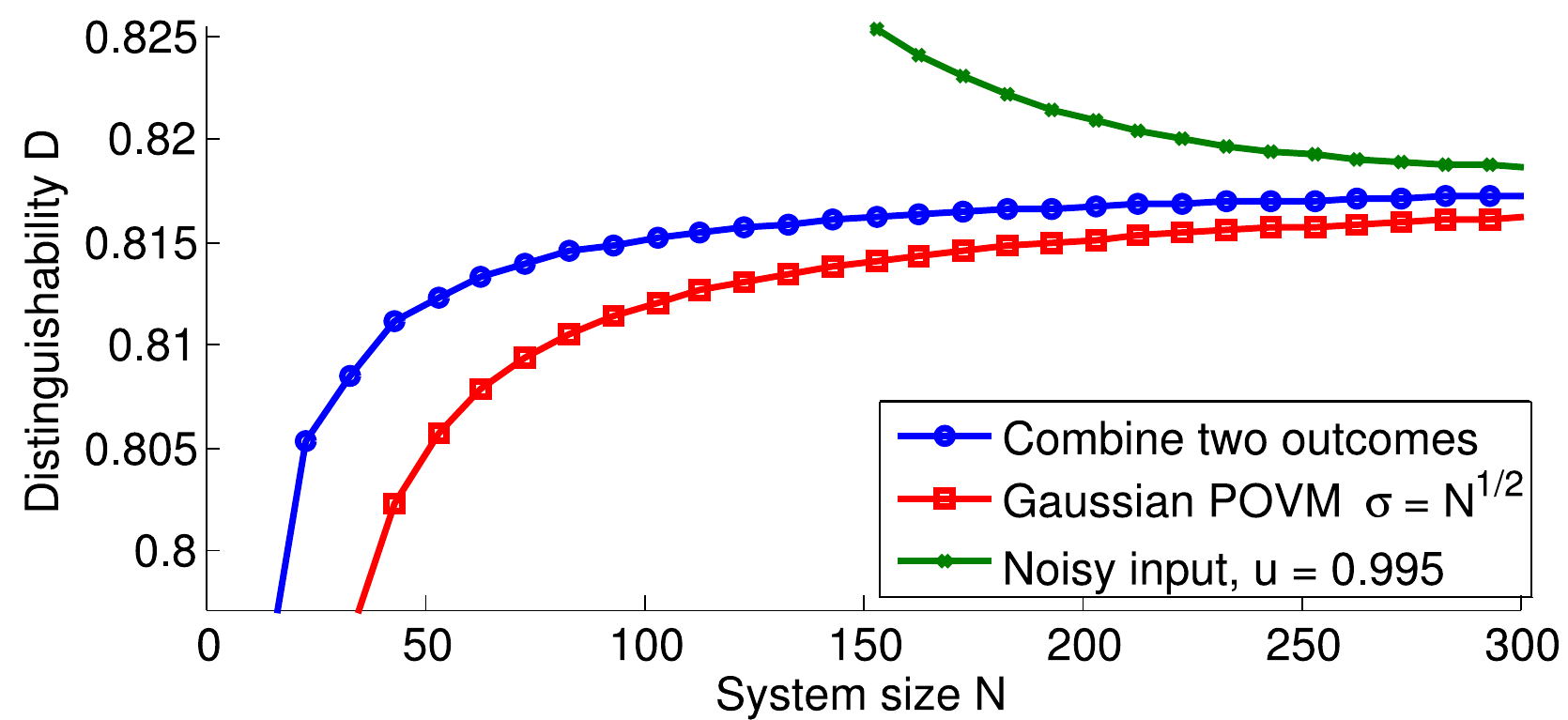}}
\caption[]{\label{fig:Dist--diffSettings}  Distinguishability $D$ for the three scenarios of imperfections discussed in the text. The plot indicates that increasing the system size does not improve the success probability to distinguish $\left| \psi^{+} \right\rangle $ from $\left| \psi^{-} \right\rangle $.}
\end{figure}

The numerical calculations are presented in more detail in the Appendix and, for a certain choice of the parameters, are plotted in Fig.~\ref{fig:Dist--diffSettings}. For all three scenarios of imperfect measurements or state preparations, a numerical extrapolation shows that the distinguishability between $\left| \psi^{+} \right\rangle $ and $\left| \psi^{-} \right\rangle $ converges to the same value $D\approx 82 \%$  \cite{NoteKorsbakken} and cannot be improved by increasing $N$.

We conclude that --according to the literature \cite{BM,KWDC,MvD}-- and arguments based on distinguishability, one cannot use optimal covariant cloning devices to generate macroscopic superpositions.

\subsection{Usefulness for parameter estimation}
In contrast to measures on macroscopic superpositions \cite{BM,KWDC,MvD} there exist also proposals for a characterization of macroscopic quantum states that do not rely on a certain structure of the state as in Eq.~(\ref{eq:2}) \cite{p-Index,LJ,FD12}. Directly applicable to spin systems are \cite{p-Index,FD12}, which qualify the cloned superposition as macroscopic.

The so-called index $p$ \cite{p-Index} --introduced by Shimizu and Miyadera-- formalizes the observation that classical probability distributions encountered in statistical mechanics show sharp peaks for local observables $M$ like the magnetization. This comes from the fact that these $M$ exhibit an expectation value $\langle M \rangle_{}\propto N$ and a variance $\mathcal{V}(M) \propto N$, which means that the relative uncertainty $\sqrt{\mathcal{V}(M)}/\langle M \rangle$ vanishes in the limit of large $N$. In contrast, there exist ``anomalously fluctuating'' pure quantum states $\left| \psi \right\rangle $ with a variance $\mathcal{V}_{\psi}(M) \propto N^2$. Such states are called macroscopic according to Ref.~\cite{p-Index}. In Ref.~\cite{FD12}, we motivate an assignment of an effective size: $N_{\mathrm{eff}}^{\mathrm{S}}(\psi) = \max_{M:local}\mathcal{V}_{\psi}(M)/N$.

The second measure that we discuss in this context is the quantum Fisher information \cite{Fisher} and is related to the work of Bj\"ork and Mana. In Ref.~\cite{FD12}, we suggest to take the usefulness for quantum applications (e.g., in terms of quantum metrology) as well as the indication of ``macroscopic quantum phenomena'' (e.g., the possibility to observe rapid oscillations generated by local Hamiltonians) as indications for macroscopic quantum states. However, we do not refer to a specific superposition as in Eq.~(\ref{eq:2}). The usefulness in quantum metrology and the possibility of rapid oscillations, respectively, can be quantified by the quantum Fisher information $\mathcal{F}$ (see Ref.~\cite{Fisher} and the Appendix and Refs.~\cite{Fleming,FD12}, respectively). Recall that for pure states $\left| \phi \right\rangle $ one has $\mathcal{F}(\phi,H)= 4 \mathcal{V}_{\phi}(H)$. Besides the different motivation, there hence exists an intimate mathematical connection between the Fisher information of pure states and the index $p$, for which reason one can assign the same effective size: $N_{\mathrm{eff}}^{\mathrm{F}}(\rho) = \max_{H:local}\mathcal{F}(\rho,H)/(4N)$ (see also Ref.~\cite{FD12}).

Already in the context of the relative Fisher information, we have calculated the maximal variance for local operators, $\mathcal{V}_{\psi}(H^{\mathrm{opt}}) = \frac{1}{2}(N+1)^2 + N +1$, leading to an effective size $N_{\mathrm{eff}}^{\mathrm{S}} = N_{\mathrm{eff}}^{\mathrm{F}} = \frac{1}{2}N + 2 + 3/(2N)$. This means that the quantum states producible by means of PCQC are  macroscopic with respect to these measures, see Table~\ref{tab:summaryMacroMeasure}.
We emphasize that the entanglement to a microscopic particle is not necessary. In fact, \textit{any} single qubit state that is subject to a PCQC results in a state that is in this sense macroscopic.

\begin{table}[htbp]
\begin{tabular}{l @{\quad}  l @{\quad} l}  \hline  \hline
Type & Authors & Effective size\\
  \hline
&Bj\"ork and Mana \cite{BM}& $1+O(1/N)$\\
MS&Korsbakken \textit{et al.} \cite{KWDC}& 2\\
&Marquardt \textit{et al.} \cite{MvD}&2\\\hline
MQ&Shimizu and Miyadera \cite{p-Index}& $\frac{1}{2}N + O(1)$\\
&Fr\"owis and D\"ur \cite{FD12}& $\frac{1}{2}N + O(1)$\\
\hline  \hline\end{tabular}
\caption[]{\label{tab:summaryMacroMeasure} Summary of the effective size due to the measures of macroscopic superpositions (MS) and macroscopic quantum states (MQ), respectively.}
\end{table}

The quantum Fisher information $\mathcal{F}$ of the pure cloned state $\left| \psi \right\rangle $ of Eq.~(\ref{eq:1}) scales with $N^2$, which is optimal for local $H$ and implies that the state can be used for quantum enhanced  metrology \cite{HeisenbergOptimal}. However, in the presence of local phase noise, the Heisenberg scaling cannot be attained anymore, and the best improvement one can gain is a constant factor $e$ compared to classical strategies \cite{LowerBoundUncertainty}. While a Greenberger-Horne-Zeilinger state in this scenario does not perform better than a product state \cite{Huelga}, spin squeezed states \cite{SSS} can asymptotically achieve this limit \cite{SqueezingMetrology}. Note that Dicke states are not spin squeezed \cite{DickeNoSSS}. Hence, the simple protocols considered in Refs.~\cite{Huelga,SqueezingMetrology} do not give an improvement compared to classical input states.

Here we study the usability of Dicke states for parameter estimation in the presence of different kinds of noise. Neglecting the micropart of Eq.~(\ref{eq:1}), we investigate the performance of the Dicke state $\left| N,\frac{1}{2}\left( N-1 \right) \right\rangle $ for a time evolution that consists of a local unitary rotation around the $x$ axis with frequency $\omega$, where in addition local bit-flip noise with a decoherence rate $\gamma$ is considered. Note that this is in contrast to Ref.~\cite{CloningMetrology}, where the use of PCQC was suggested to amplify the signal \textit{after} the unitary time evolution. We consider two scenarios: optimal local and global measurements. As in Refs.~\cite{Huelga,SqueezingMetrology} and other references, we give the relative improvement of those scenarios compared to the best classical protocol. We find that the optimal global measurement leads to a better performance than the classical strategies, similarly as the spin squeezed states. However, within the numerical optimization procedure, we were not able to find a symmetric local measurement that could beat the classical protocol. This means that while full control over the measurement setup directs to higher sensitivity, this improvement is lost if only local measurements are available. Qualitatively, the same conclusion can be drawn for white noise as a decoherence source, which is --in addition to illustrations and further details-- discussed in the Appendix.

\subsection{Conclusion and outlook}
We have discussed the generation of macroscopic quantum states by means of optimal phase covariant cloning. We have argued that neither with proposals from the literature nor by arguments on the distinguishability it is justified to call the generated micro-macro state (\ref{eq:1}) a macroscopic quantum superposition in the spirit of Schr\"odinger's cat.
It is important to note that these results do not depend on the specific mapping of Eq.~(\ref{eq:3}). They are valid in general for optimal phase covariant cloning maps --and hence apply also to the experiments of Refs.~\cite{DeMartini08,ExpRealizationPCQC}-- because of the restrictions imposed by the no-cloning theorem. The inability of perfect cloning leads to a reduction of the effective sizes according to Refs.~\cite{BM,KWDC,MvD,FD12} and to a decreased distinguishability.

On the other hand, we have seen that this state is indeed a macroscopic quantum state since it can show nonclassical behavior, in particular an increased sensitivity in quantum metrology even in the presence of noise. Our results indicate that a careful investigation is required to judge whether large-scale quantum systems can indeed be considered to be truly macroscopic. While we have concentrated here on systems consisting of many spins, similar investigations, e.g., for superposition states produced in optomechanical systems are highly interesting, given the spectacular experimental progress in this area.

\textit{Acknowledgments.---} The research was funded by the Austrian Science Fund (FWF):  P20748-N16, P24273-N16, SFB F40-FoQus F4012-N16 and the European Union (NAMEQUAM).

\appendix

\subsection{Measuring a subgroup}
To determine the effective size of the macro-measure of Korsbakken and coworkers \cite{KWDC}, we calculate the distinguishability of $\left| \psi^{+} \right\rangle $ and $\left| \psi^{-} \right\rangle $ if we only measure $k$ out of $N$ particles (see also Ref.~\cite{De12}). The figure of merit is the success probability $P = \frac{1}{2} + \frac{1}{4} \lVert \rho_k^+ - \rho_k^{-} \rVert_1$ with $\rho_k^{\pm} = \mathrm{Tr}_{N \setminus k}\left| \psi^{\pm} \rangle\!\langle \psi^{\pm}\right| $. We use the general decomposition
\begin{equation}
\ket{N,x} = \sum_{j = 0}^k\sqrt{\frac{\binom{k}{j}\binom{N-k}{x-j}}{\binom{N}{x}}}  \ket{k,j} \otimes \ket{N-k,x-j},\label{eq:5}
\end{equation}
where $\binom{N}{x}$ is the binomial coefficient that gives the number of possible permutations.
Defining $c_j^{x} = \binom{k}{j}\binom{N-k}{x-j}/\binom{N}{x}$, we get
\begin{equation}
  \begin{split}
    \rho_k^{\pm} &= \frac{1}{2}\sum_{j=0}^k
    (c_j^{\frac{N-1}{2}}+c_j^{\frac{N+1}{2}}) \ketbrad{k,j} \\ & \pm
    \frac{1}{2}\sum_{j = 0}^{k-1} \sqrt{c_j^{\frac{N-1}{2}}c_{j+1}^{\frac{N+1}{2}}}\,\left(\ketbra{k,j}{k,j+1}+ \mathrm{h.c.}\right)\label{eq:4}
  \end{split}
\end{equation}
We consider in the following large particle numbers $N$. For $x\in \{ \frac{1}{2}(N-1),\frac{1}{2}(N+1) \}$ we can approximate $\binom{N-k}{x-j} /\binom{N}{x} = 2^{-k}[1+O(k/N)]$. With $\sqrt{c_j^{\frac{N-1}{2}}c_{j+1}^{\frac{N+1}{2}}} = c_j^{\frac{N-1}{2}}\sqrt{(k-j)/(j+1)}$, the trace norm of
\[\rho_k^{+}-\rho_k^{-} \approx \frac{1}{2^k}\sum_{j=0}^{k-1}\binom{k}{j}\sqrt{\frac{k-j}{j+1}}
\left(\ketbra{k,j}{k,j+1}+ \mathrm{h.c.}\right)  \]
can be easily computed numerically. From Eq.~(\ref{eq:5}), the same calculations can be done to find $\rho_{N-k}^{\pm}$ in the limit of large $N$. The results are discussed in the main text.

\subsection{Maximal variance of micro-macro state}
Given the $N+1$ particle state $\left| \phi \right\rangle $, we search for the local operator $M = \sum_{i=1}^{N+1}\sum_{j=x,y,z} \alpha_{i,j}\sigma_j^{(i)}$ with maximal variance $\mathcal{V}_{\phi}(M)$, under the condition that $\sum_{i=1}^{N+1}\sum_{j=x,y,z} \alpha_{i,j}^2 = N+1$. We need this in order to decide on the effective size of several measures \cite{p-Index,BM,FD12}. The covariance-matrix method \cite{Covariance} is an efficient method to find the optimal set $\{\alpha_{i,j}\}$. With $\Delta\sigma_j^{(i)} = \sigma_j^{(i)} - \langle \sigma_j^{(i)} \rangle_{\psi}$ we define the hermitian covariance-matrix [$(i,j)$ now serve as a multi-index $C_{(i,j),(i\prime,j\prime)}^{\psi} = \langle \Delta\sigma_j^{(i)}  \Delta\sigma_{j\prime}^{(i\prime)}\rangle_{\psi}$]. The maximal eigenvalue of $C$ is the maximal variance of a local observable $M$ and entries of the corresponding eigenvector are --up to a normalization-- the weights $\alpha_{i,j}$.

For $\left| \psi \right\rangle $ in Eq.~(\ref{eq:1}) of the main text, the dimension of $C$ is even more reduced, since the macro-part of $\left| \psi \right\rangle $ with $N$ particles is symmetric under particle permutations and therefore we can make the ansatz $M = \sum_{j=x,y,z}\alpha_{j,1}\sigma_j^{(1)} + \alpha_{j,2}\sum_{i=2}^{N+1}\sigma_j^{(i)}$. The normalization reads $\sum_{j=x,y,z}\alpha_{j,1}^2 + \alpha_{j,2}^2 N = N + 1$. The calculation of $C$ renders easily and we get
\begin{equation}
\label{eq:7}
C^{\psi}\!\!=\!\!\begin{pmatrix}
  1&0&0&-\frac{N+1}{2}&0&0\\
  0&1&0&0&-\frac{N+1}{2}&0\\
  0&0&1&0&0&0\\
  -\frac{N+1}{2}&0&0&\frac{N^2+2N-1}{2}&0&0\\
  0&-\frac{N+1}{2}&0&0&\frac{N^2+2N-1}{2}&0\\
  0&0&0&0&0&1\\
\end{pmatrix}\!\!.
\end{equation}
The matrix consists of three independent $2\times 2$-blocks: an identity matrix and two times the matrix
\begin{equation}
c=\begin{pmatrix}
  1&-\frac{N+1}{2}\\
  -\frac{N+1}{2}&\frac{N^2+2N-1}{2}\\
\end{pmatrix},\label{eq:8}
\end{equation}
which belong to the coefficient pairs $\alpha_x\equiv(\alpha_{1,x},\alpha_{2,x})$ and $(\alpha_{1,y},\alpha_{2,y})$, respectively. We search for the maximal expectation value $\alpha_x c \alpha_x^T = \alpha_{1,x}^2 + \alpha_{2,x}^2 \left[ \frac{\left( N+1 \right)^2}{2}-1 \right]-\alpha_{1,x}\alpha_{2,x}\left( N+1 \right)$. After inserting the normalization condition $\left| \alpha_{1,x} \right|=\sqrt{N\left( 1-\alpha_{2,x} \right)+1}$ (and fixing $\alpha_{1,x}$ to be positive), we find that $\alpha_x = (1,-1)$ is an optimal choice. This leads to a maximal variance $\mathcal{V}_{\psi}(M^{\mathrm{opt}}) = \frac{1}{2}(N+1)^2 + N +1$. Notice that any collective local rotation around the $z$-axis $e^{(-i\alpha \sigma_z)\otimes (N+1)} M^{\mathrm{opt}}\, e^{(i\alpha \sigma_z) \otimes (N+1)} , \alpha \in \mathbbm{R}$ gives the same variance.

In addition, we search for the optimal measurement for the $\left| \phi_0 \right\rangle = |+\rangle\otimes |\psi^{-}\rangle$ state. Since it is a product state between the first particle and the rest, the variances of both parts sum up $\mathcal{V}_{\phi_0}(M) = \mathcal{V}_{\left| + \right\rangle }(M) + \mathcal{V}_{\psi_{-}}(M)$. The micro-part has with $\sigma_y$ a maximal variance equaling one, while for the macro-part one can use once again the covariance method for the measurement $M = \sum_{j=x,y,z} \alpha_{j,2}\sum_{i=2}^{N+1}\sigma_j^{(i)}$. This leads to the matrix
\begin{equation}
\label{eq:9}
C^{\psi^{-}} =
\begin{pmatrix}
  \frac{N^2-1}{4}&0&0\\
  0&\frac{N^2+2N-1}{2}&0\\
  0&0&1\\
\end{pmatrix}
\end{equation}
and shows that the maximal variance is governed by a measurement in $y$-direction, which leads to $\mathcal{V}_{\phi_0}(M^{\mathrm{opt}}) = 1 + \frac{1}{2}(N^2+2N-1) = \frac{1}{2}(N+1)^2$.

\subsection{Details on the distinguishability} This paragraph contains some details of the calculation concerning the distinguishability of $\left| \psi^{+} \right\rangle $ and $\left| \psi^{-} \right\rangle $ under imperfect conditions. We therefore calculate the probabilities $\langle \pi_i \rangle_{\psi^{\pm}}$ which are used in the first scenario in the main text. Using Eq.~(\ref{eq:5}), the states $\left| \psi^{\pm} \right\rangle $ can be expressed as superposition of Dicke states $\left| N,k \right\rangle_x = \mathit{Had}^{\otimes N} \left| N,k \right\rangle$ in the $x$-basis ($\mathit{Had}$ denoting the Hadamard operator)
\begin{equation}
\label{eq:11}
\left| \psi^{\pm} \right\rangle  = \sum_{k = 0}^N\frac{1 \pm (-1)^k}{\sqrt{2}} \sqrt{\binom{N}{k}} \beta_k  \left| N,k \right\rangle_x
\end{equation}
with $\beta_k = 1/\sqrt{2^{N-1}\binom{N}{\frac{N-1}{2}}} \sum_{i = 0}^N (-1)^i \binom{k}{i}\binom{N-k}{\frac{N-1}{2}-i}$. With this we easily see that
\begin{equation}
\label{eq:14}
\langle \pi_i \rangle_{\psi^{\pm}} =  \frac{\left[1 \pm (-1)^k\right]^2}{2} \binom{N}{i} \beta_i^2.
\end{equation}

\begin{figure}[htbp]
\centerline{\includegraphics[width=.8\columnwidth]{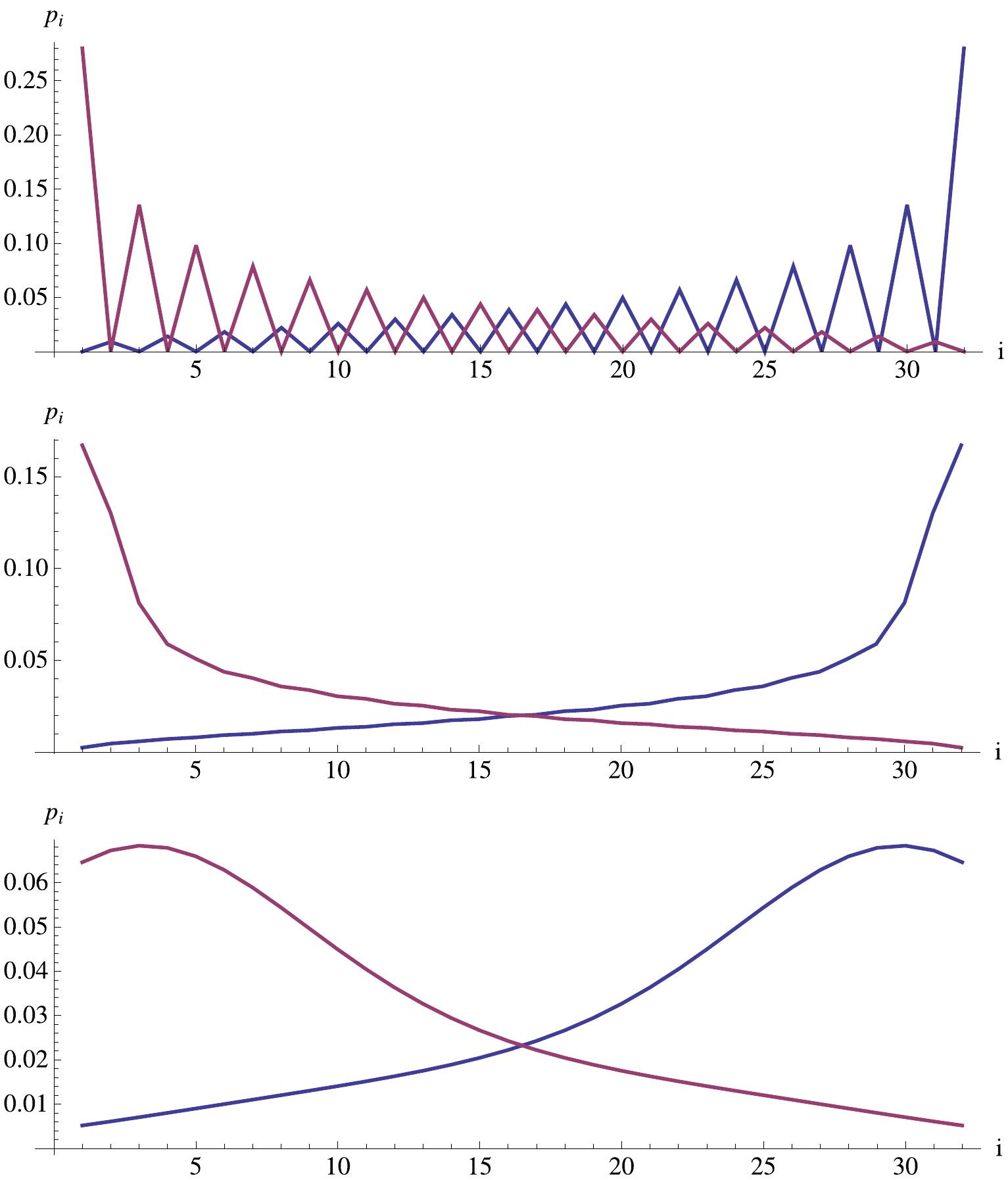}}
\caption[]{\label{fig:distinguishability}Probabilities $p_i=\langle E^2_i \rangle_{\psi^{\pm}}$ are plotted for $N = 31$ and three different choices of $\sigma$: 0 [i.e., perfect measurement Eq.~(\ref{eq:14})], 1 and $\sqrt{N}$ (from top to bottom). Already a uncertainty of $\sigma= 1$ reduces the distinguishability between $\left| \psi^{+} \right\rangle $ (blue) and $\left| \psi^{-} \right\rangle $ (red) drastically.}
\end{figure}

In the same manner, one can calculate the probabilities under the assumption of a POVM with Kraus-operators $E_j$ defined in the main text. We arrive at
\begin{equation}
\label{eq:13}
\langle E^2_i \rangle_{\psi^{\pm}} = \sum_{k = 0}^N \frac{\left[1 \pm (-1)^k\right]^2}{2} \binom{N}{k} \beta_k^2 n_k e^{-\frac{(k-i)^2}{2\sigma^2}}.
\end{equation}
A numerical example for $\langle E^2_i \rangle_{\psi^{\pm}}$ is shown in Fig.~\ref{fig:distinguishability}.

Exchanging the pure input states $\left| \psi^{\pm} \right\rangle $ by noisy $\mathcal{E}_z \left( \left| \psi^{\pm} \rangle\!\langle \psi^{\pm}\right|  \right)$ is more involved. An error on a spin particle leads to a wrong assignment of the measurement outcome, similarly as with the POVM measurement, but multiple errors can cancel this effect. Considering carefully all posibilities, we end up with the formula

\begin{equation}
\label{eq:12}
\begin{split}
  p_i^{\pm} &= \mathrm{Tr}\left[ \pi_i \mathcal{E}_z \left( \left|
        \psi^{\pm} \rangle\!\langle\psi^{\pm} \right| \right) \right]\\
  &= \sum_{k = 0}^N \frac{\left[1 \pm (-1)^k\right]^2}{2} \binom{N}{k} \beta_k^2 \eta_k
\end{split}
\end{equation} with
\begin{equation}
\label{eq:15}
\eta_k=\sum_{l = \frac{\left| i-k \right|}{2}}^{N/2}u^{N-2l}\left( 1-u \right)^{2l}\!\binom{k}{l+\frac{k-i}{2}}\!\binom{N-k}{l-\frac{k-i}{2}}.
\end{equation}

Fig.~\ref{fig:distinguishability} nicely illustrates the discussion in the main text. The distinction can be done on two different scales: on the ``microscopic'' scale, where we can perfectly distinguish between $\left| \psi^{\pm} \right\rangle $ measuring the parity, and on the ``macroscopic'' scale, where one tries to compare the total number of excitations. The micro-level is very fragile against any kind of disturbance and the distinguishability is lost almost instantaneously. The more robust distinction on the macro-level suffers from the large variances of the observable and one therefore cannot distinguish with arbitrary precision.
\subsection{Parameter estimation protocol}
 Here we give some details on the time evolution and the measurements to calculate the so-called Cram\'er-Rao bound \cite{Fisher} for the quantum state $\left| N,\frac{1}{2}(N-1) \right\rangle $. The Cram\'er-Rao bound is a lower bound on the minimal error $\delta\omega$ one has to assign for the estimation of a parameter $\omega$. The time evolution that we consider is governed by the master equation
\begin{equation}
\dot{\rho}(t) = i \frac{\omega}{2} \left[ \rho,\sum_{i=1}^{N}\sigma_x^{(i)} \right] + \frac{\gamma}{2} \sum_{i = 1}^N \left(\sigma_x^{(i)} \rho \sigma_x^{(i)} - \rho\right)\label{eq:17}
\end{equation}
with the parameters $\omega > 0$ for the unitary rotation and $\gamma >0$ for the decoherence rate. The quantum state under this evolution is $\rho(t) = e^{\left( -i \omega t/2 \right)\otimes N} \mathcal{E}_x \left( \left|N,\frac{N-1}{2}  \rangle\!\langle N,\frac{N-1}{2}\right|  \right) e^{\left( i \omega t/2 \right)\otimes N}$, recalling that $\mathcal{E}_x(\rho) = \prod_{i=1}^{N} u \rho + (1-u) \sigma_x \rho \sigma_x$ with $u = \frac{1}{2}\left( 1+e^{-\gamma t} \right) \in [\frac{1}{2},1]$.

We calculate the Fisher information of $\rho(t)$ for the optimal measurement --the so-called quantum Fisher information $\mathcal{F}$-- and a specific local measurement $M_j = \sum_{i= 1}^N\sigma_j^{(i)}$. The latter is also referred to as classical Fisher information $F$. Here, $\mathcal{F}$ is given by \cite{Fisher}
\begin{equation}
\label{eq:6}
\mathcal{F}(\rho,H) = \sum_{i,j}2 \frac{\left[ p_i(t)-p_j(t) \right]^2}{p_i(t)+p_j(t)} \left| \left\langle i \right| t H/2 \left| j \right\rangle  \right|^2,
\end{equation}
where $p_j(t)$ are the eigenvalues of the density matrix $\rho(t)$, that is, the evolving state without unitary rotation. The corresponding eigenvectors $\left| j \right\rangle$ are time independent. We calculate the spectral decomposition of $\rho(t)|_{\omega=0}$ numerically and are therefore restricted to rather small system sizes.

\begin{figure}[htbp]
\centerline{\includegraphics[width=.9\columnwidth]{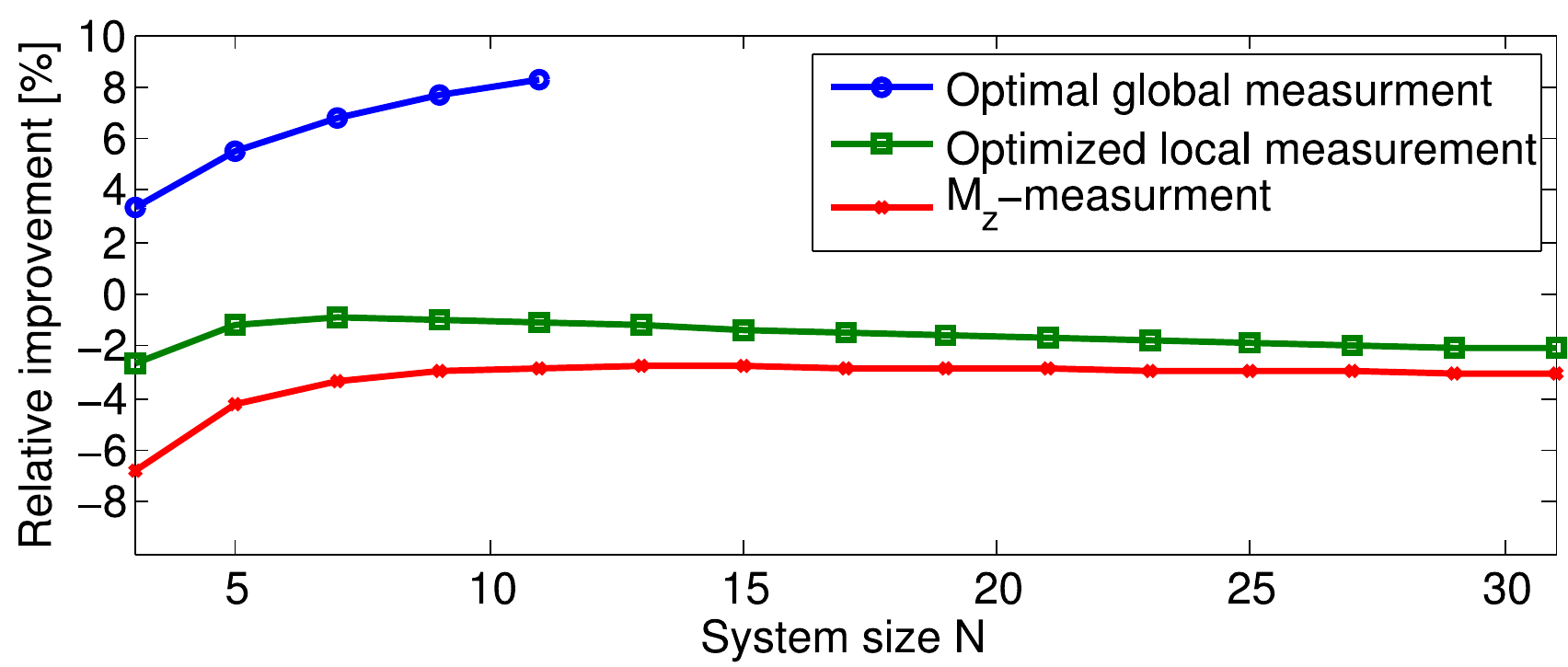}}
\caption[]{\label{fig:PhaseEst} Relative improvement of phase estimation protocols using $\left| N,\frac{1}{2}\left(N-1  \right) \right\rangle $ as a starting state with the parameters $\omega = 1$ and $\gamma = 0.5$. We compare the three scenarios global and local optimal measurement as well as fixed measurement in $z$-direction to the best strategy with product states as inputs. The global measurement can lead to an increased sensitivity for the investigated system sizes. Note that the maximal possible improvement is 40\% \cite{LowerBoundUncertainty}. On the other hand, local measurement strategies fail to overcome classical protocols. The different ranges of data points results from different numerical techniques (see text).}
\end{figure}

\begin{figure}[htbp]
\centerline{\includegraphics[width=.9\columnwidth]{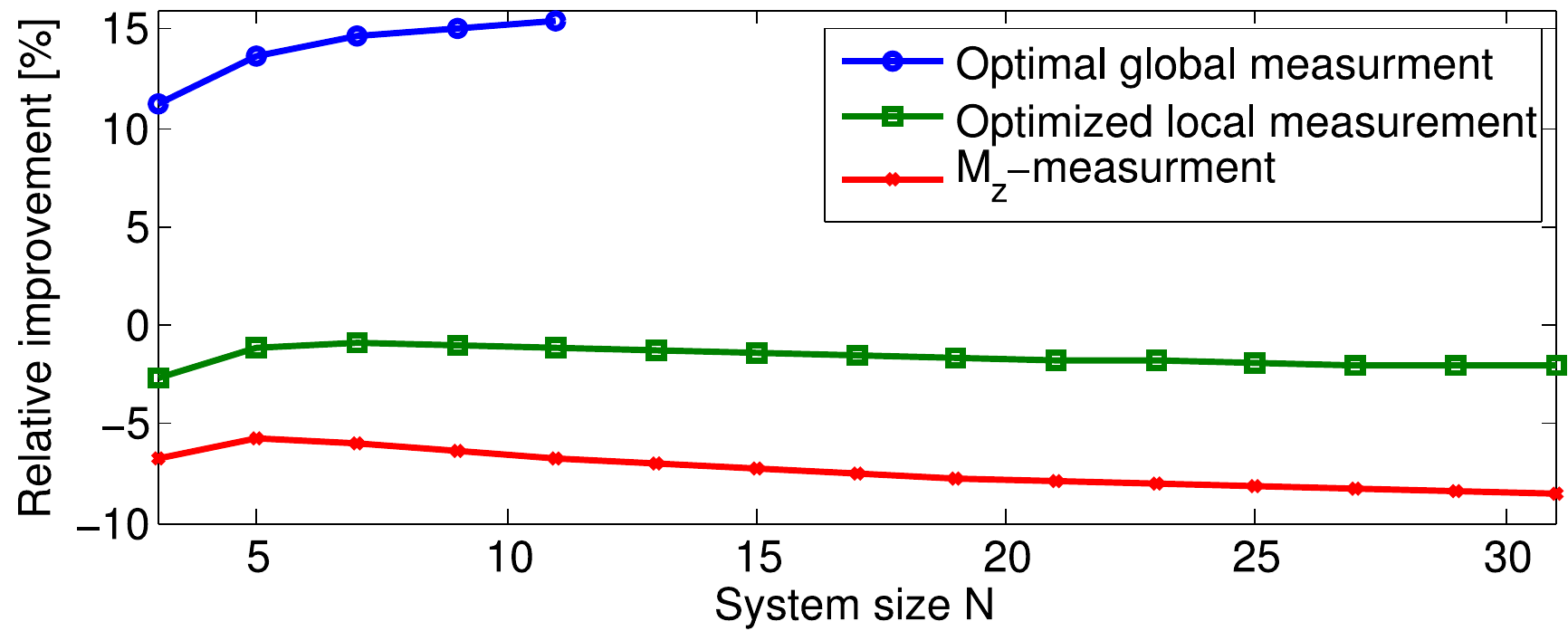}}
\caption[]{\label{fig:PhaseEstWN} With white noise as decoherence source, the uncertainty of the estimation protocol compared to classical strategy is the same for our choice of local measurements, while the relative improvement changes for the optimal global measurement (see also Fig.~\ref{fig:PhaseEst}); parameters: $\omega = 1$ and $\gamma = 0.2$.}
\end{figure}

The the Cram\'er-Rao bound in this scenario reads $\delta \omega = 1/\sqrt{n \mathcal{F}}$, where $n$ is the total number of repetitions. Assuming a total time $T$ for the experiment, we have $n = T/t$. We then minimize $\delta\omega$ with respect to $t$. We compare $\delta\omega$ to product states as input states. The same procedure leads there to the uncertainty $\delta\omega_{\mathrm{PS}} = \sqrt{2e\gamma/(T N)}$ \cite{Huelga}. We are interested in  the relative improvement $1-\delta\omega/\delta\omega_{\mathrm{PS}}$, which is the quantity plotted in Figs.~\ref{fig:PhaseEst} and \ref{fig:PhaseEstWN}.

Similarly, we proceed with the classical Fisher information $F$. For a discrete probability distribution $s_i(\omega,t)$ that depends on the parameter $\omega$, $F$ is defined by \cite{ClassicalFisher}
\begin{equation}
\label{eq:10}
F = \sum_{i} s_i \left(  \frac{d \log s_i}{d \omega}\right)^{2}.
\end{equation}
Here, $s_i(\omega,t)$ are probabilities of the results for the different outcomes of a given measurement $M_j$. We start with the measurement $M_z=\sum_{i = 1}^N \sigma_z^{(i)}$, where we can find analytical expressions for $s_i(\omega,t) = \langle \pi_i \rangle_{\rho(t)}$, where $\pi_i$ is the projector on the subspace spanned by all product states with $i$ times $\left| 1 \right\rangle  $ states and $N-i$ times $\left| 0 \right\rangle $ states. This is done by using the splitting in Eq.~(\ref{eq:5}) for $\left| N,\frac{1}{2}(N-1) \rangle\!\langle N,\frac{1}{2}(N-1) \right| $ with respect to $i:N-i$ and a consequent expression of the sub-parts in the $x$-basis, similar to Eq.~(\ref{eq:11}). We find
\begin{equation}
\label{eq:16}
\begin{split}
  &s_i(\omega,t) = \langle \pi_i \rangle_{\rho(t)}\\ &=
  \binom{N}{i} \left\langle 1 ^{\otimes i}\otimes  0 ^{\otimes N-i}\right|  \rho(t)\left| 1 ^{\otimes i}\otimes  0 ^{\otimes N-i} \right\rangle \\ &
 = \sum_{j,j\prime = J}\mu_j\,\mu_{j\prime}\,\Gamma_{j,j\prime}^{i,-}\,\Gamma_{\frac{N-1}{2}-j,\frac{N-1}{2}-j\prime}^{N-i,+}
\end{split}
\end{equation}
with the definitions
\begin{equation}
\label{eq:19}
\begin{split}   \mu_{j}&=\sqrt{\frac{\binom{N}{i}\binom{i}{j}\binom{N-i}{\frac{N-1}{2}-j}}{\binom{N}{\frac{N-1}{2}}}},\\
   \Gamma_{j,j\prime}^{i,\pm} &= \sum_{l,l\prime,m = 0}^i (\pm
  1)^{l+l\prime}
  \nu_{j,l}^i\,\nu_{j\prime,l\prime}^i\,\nu_{m,l}^i\,\nu_{m,l\prime}^i \times \\ & \qquad u^{i-m}u^m e^{-i\omega t(l-l\prime)},\\
   \nu_{j,l}^i &= \sqrt{\frac{\binom{i}{l}}{2^i\binom{i}{j}}}\sum_{m = M}(-1)^m\binom{l}{m}\binom{i-l}{j-m}.
\end{split}
\end{equation}
In Eq.~(\ref{eq:16}) we have $J = \left\{ \max \left( 0,i-\frac{1}{2}(N+1) \right),\ldots, \min \left(\frac{1}{2}(N+1),i  \right) \right\}$ and in Eq.~(\ref{eq:19}) $M=\left\{ \max \left( 0,j-i+l \right),\dots, \min \left(l,j \right) \right\}$.

From this we can calculate $F$ and pursue as before with $\mathcal{F}$. To improve the results, we can try to optimize the local measurement. We test local measurements where every qubit is measured
in the same basis. Furthermore, we assume to find the best basis in the $y-z$-plane of the Bloch sphere of a single spin, because measuring in $x$-direction only gives time-independent results. Instead of $\left| 0 \right\rangle $ and $\left| 1 \right\rangle $, we now project onto $e^{-i\alpha \sigma_x}\left| 0 \right\rangle $ and $e^{-i\alpha \sigma_x}\left| 1 \right\rangle $, $\alpha \in \left[ 0,\pi/2 \right]$. This can be easily implemented in Eqs.~(\ref{eq:16}) and (\ref{eq:19}), exchanging the term $e^{-i\omega t(l-l\prime)} $ by $ e^{-i(\omega t +\alpha)(l-l\prime)}$. We search numerically for the optimal $\alpha$ and present the results in Fig.~\ref{fig:PhaseEst}. We find indeed an improvement with respect to $M_z$, which is however still above the uncertainty $\delta\omega_{\mathrm{PS}}$. Whether a more general, asymmetric measurement can improve the sensitivity is an open question.

The very same can be exercised with different noise channels, as long as it commutes with the unitary evolution in $x$-direction. An example is the so-called white noise. This channel is defined as $\mathcal{E}_{\mathrm{W}}(\rho) = \prod_{i=1}^N \left(p \rho +\frac{1-p}{4} \sum_{i = 0}^3 \sigma_i\rho\sigma_i \right)$ with $p=e^{-\gamma t}$. The results for the quantum Fisher information (\ref{eq:6}) are slightly better compared to the phase-error channel $\mathcal{E}_x$, while the formulas Eqs.~(\ref{eq:16}) and (\ref{eq:19}) are indeed also valid for white noise if we identify $u = (1+p)/2$ (see Fig.~\ref{fig:PhaseEstWN}).

\begin{thebibliography}{99}
\bibitem[]{NoteNotation}  The notation $\sigma_x^{(i)}$ means that $\sigma_x$ acts on the $i^{\mathit{th}}$ particle.
\bibitem[]{NoteBM} We assume that the quantum states $\left| \phi_0 \right\rangle $ and $\left| \phi_1 \right\rangle $ exhibit the same values for the orthogonalization time and variance, which is valid here.
\bibitem[]{NoteKorsbakken} This value is numerically very similar to the distinguishability of $\left| \phi_0 \right\rangle $ and $\left| \phi_1 \right\rangle $ if we measure $1\ll k \ll N$ particles, as we find for the macro-measure of Korsbakken. This indicates that the loss of particles leads to the same ability of distinction as the phase-noise channel.
\end{thebibliography}
\end{document}